# Language ID Prediction from Speech Using Self-Attentive Pooling and 1D-Convolutions

**Roman Bedyakin[1], Nikolay Mikhaylovskiy[2,3]**

[1]AO HTSTS, Moscow, Russia, [2]NTR Labs, Moscow, Russia, [3]Tomsk State University, Tomsk, Russia

{rbedyakin, nickm}@ntr.ai

## Abstract

This memo describes NTR-TSU submission for SIGTYP 2021 Shared Task on predicting language IDs from speech.

Spoken Language Identification (LID) is an important step in a multilingual Automated Speech Recognition (ASR) system pipeline. For many low-resource and endangered languages, only single-speaker recordings may be available, demanding a need for domain and speaker-invariant language ID systems. In this memo, we show that a convolutional neural network with a Self-Attentive Pooling layer shows promising results for the language identification task.

## 1 Introduction

Spoken Language Identification (LID) is a process of classifying the language spoken in a speech recording and is an important step in a multilingual Automated Speech Recognition (ASR) system pipeline.

Differences between languages exist at all linguistic levels and vary from marked, easily identifiable distinctions (such as the use of entirely different words) to more subtle variations, which might have been lost or gained due to language contact. The latter end of the range is a challenge not only for automatic LID systems but also for linguistic sciences themselves.

In this memo, we show that a convolutional neural network with a Self-Attentive Pooling layer shows promising results in low-resource setting for the language identification task. The system described herein is identical to the one simultaneously submitted for Low Resource ASR challenge at Dialog2021 conference, language identification track, although the dataset is completely different.

### 1.1 Previous work

The first works on LID date back at least to mid-seventies, when Leonard and Doddington (1974) explored frequency of occurrences of certain reference sound units in different languages.

Previously developed LID approaches include:

- Purely acoustic LID that aims at capturing the essential differences between languages by modeling distributions in a compact representation of the raw speech signal directly.

- Phonotactics LID rely on the relative frequencies of sound units (phoneme/phone) and their sequences in speech.

- Prosodic LID use tone, intonation and prominence, typically represented as pitch contour.

- Word Level LID systems use fully-fledged large vocabulary continuous speech recognizers (LVCSR) to decode an incoming utterance into strings of words and then use Written Language Identification.

In the latest 10 years, intermediary-dimensional vector representations similar to i-vector (Dehak, et al. 2011a, 2011b, Kanagasundaram et al., 2011) and x-vector (Snyder et al., 2018) have been dominating the speech classification field, including LID. Additionally, starting from 2014 (Lopez-Moreno et al., 2014), deep neural networks have been predominantly used for such tasks (see, for example, Bartz et al, 2017), Abdullah et al., 2020), Draghici et al, 2020), van der Merwe, 2020).

## 2  Model architecture

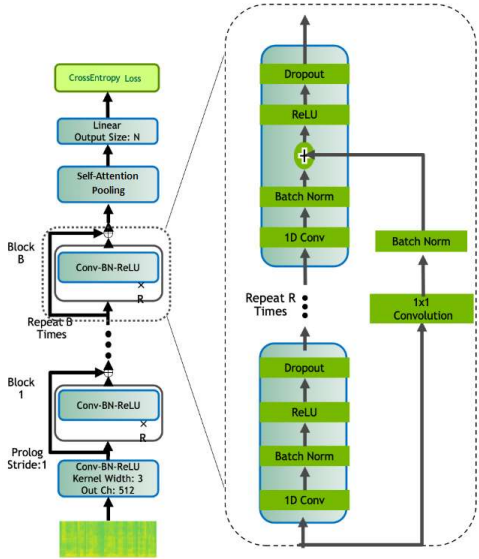

Figure 1: The model architecture

Similar to work of Koluguri et al., 2020, the model is based on 1D convolutions, namely, the QuartzNet ASR architecture (Kriman et al., 2020) comprising of an encoder and decoder structures.

### 2.1  Encoder

The encoder used is QuartzNet BxR model shown in Figure 1, and has B blocks, each with R sub-blocks (Kriman et al., 2020).

The first block is fed with MFSC coefficients vector of length 40. Each sub-block applies the following operations (Kriman et al., 2020):

- a 1D convolution,
- batch norm,
- ReLU, and
- dropout.

All sub-blocks in a block have the same number of output channels. These blocks are connected with residual connections (Kriman et al., 2020).

We use QuartzNet 15*5, with 512 channels. All the convolutional layers have stride 1 and dilation 1 (Kriman et al., 2020).

### 2.2  Self-attentive pooling decoder

Similar to Cai et al., 2018, Chowdhury et al., 2018, we agree that not all frames contribute equally to the utterance level representation. Thus we use a self-attentive pooling (SAP) layer introduced by Cai et al., 2018 to pay more attention to the frames that are more important.

Namely, we first feed the frame level feature maps $\{x_1, x_2, \cdots, x_L\}$ into a fully-connected layer to get a hidden representation

$$h_t = tanh(Wx_t + b)$$

Then we measure the importance of each frame as the similarity of $h_t$ with a learnable context vector μ and get a normalized importance weight $w_t$ through a softmax function (Cai et al., 2018).

After that, the utterance level representation $e$ can be generated as a weighted sum of the frame level feature maps based on the learned weights:

$$e = \sum_{t=1}^{T} w_t x_t$$

### 2.3  Loss Function

We have used cross-entropy loss function for this task.

## 3  Experiments

### 3.1  Datasets and tasks

For training models, speech data from the CMU Wilderness Dataset (Black, 2019) were used, which contain read speech from the Bible in 699 languages, but usually recorded from a single speaker. This training data were released in the form of derived MFCCs. The evaluation (validation, test) data come from different sources, in particular data from the Common Voice project, several OpenSLR corpora (SLR24 (Juan et al., 2014a, 2014b), SLR35, SLR36 (Kjartansson et al,. 2018), SLR64, SLR66, SLR79 (He et al., 2020), and the Paradisec collection.

There are 16 languages in the released train data, 4000 utterances per language. Table 1 summarizes the languages in the dataset.

Validation and test data consist of 8000 utterances, 500 for each language.

Table 1: Summary of languages in the dataset

| ISO 639-3 code | Language name | Genus | Family |
|---|---|---|---|
| kab | Kabyle | Berber | Afro-Asiatic |
| ind | Indonesian | Malayo-Sumbawan | Austronesian |
| sun | Sundanese | Malayo-Sumbawan | Austronesian |
| jav | Javanese | Javanese | Austronesian |
| eus | Euskara | Basque | Basque |
| tam | Tamil | Southern Dravidian | Dravidian |
| kan | Kannada | Southern Dravidian | Dravidian |
| tel | Telugu | South-Central Dravidian | Dravidian |
| hin | Hindi | Indic | Indo-European |
| por | Portuguese | Romance | Indo-European |
| rus | Russian | Slavic | Indo-European |
| eng | English | Germanic | Indo-European |
| mar | Marathi | Indic | Indo-European |
| tha | Thai | Kam-Tai | Tai-Kadai |
| iba | Iban | Malayo-Sumbawan | Austronesian |
| cnh | Chin, Hakha | Gur | Niger-Congo |

## 3.2 Optimization and training process

We have used the attention vector size of 256.

Models were trained until they reached a plateau on a validation set. Training was done using the Stochastic Gradient Descent optimizer with initial learning rate of 0.005 and cosine annealing decay to 1e-4.

## 4 Results and Discussion

We have experimented with SpecAugment augmentation introduced by Park et al., 2019 and run experiments both with and without augmentation.

The system described above allowed us to achieve the following results on the validation set (see Table 2). Somewhat surprisingly, detection of most languages was better without SpecAugment, Sundanese, Portuguese, Russian, and Iban being exceptions. Iban did not detect at all without augmentation. We can hypothesize that SpecAugment is more favorable for Indo-European and Austronesian languages detection than for other language families. This hypothesis requires further research.

Looking at the confusion matrix (Figure 2) we can see that most language samples determined as English, Kabyle or Telugu, independent of the language family. This means that there are more prominent speech features that hinder the language identification. Given the nature of the training set, that may be related to the gender of the readers.

Table 2: Results on validation dataset

| | | Without augmentation | | | With augmentation | | |
|---|---|---|---|---|---|---|---|
| Language | support | precision | recall | f1-score | precision | recall | f1-score |
| kab | 500 | **0.0735** | **0.218** | **0.11** | 0.0675 | 0.206 | 0.1017 |
| ind | 500 | 0.1102 | **0.13** | **0.1193** | **0.125** | 0.078 | 0.0961 |
| sun | 500 | 0.0747 | 0.082 | 0.0782 | **0.0753** | **0.1** | **0.0859** |
| jav | 500 | **0.0692** | 0.054 | **0.0607** | 0.0624 | **0.056** | 0.059 |
| eus | 500 | **0.1925** | **0.072** | **0.1048** | 0.1656 | 0.05 | 0.0768 |
| tam | 500 | **0.3108** | **0.304** | **0.3074** | 0.2244 | 0.14 | 0.1724 |
| kan | 500 | **0.0339** | **0.004** | **0.0072** | 0.0149 | 0.002 | 0.0035 |
| tel | 500 | **0.0298** | 0.112 | **0.0471** | 0.0284 | **0.12** | 0.0459 |
| hin | 500 | **0.0933** | **0.014** | **0.0243** | 0.0896 | 0.012 | 0.0212 |
| por | 500 | 0.0871 | 0.062 | 0.0724 | **0.1061** | **0.098** | **0.1019** |
| rus | 500 | 0.0482 | 0.032 | 0.0385 | **0.0712** | **0.038** | **0.0495** |
| eng | 500 | **0.2065** | 0.406 | **0.2738** | 0.1972 | **0.428** | 0.27 |
| mar | 500 | 0.3491 | **0.118** | **0.1764** | **0.3654** | 0.076 | 0.1258 |
| tha | 500 | **0.2167** | **0.026** | **0.0464** | 0.1014 | 0.014 | 0.0246 |
| iba | 500 | 0 | 0 | 0 | **0.0638** | **0.012** | **0.0202** |
| cnh | 500 | **0.2039** | **0.104** | **0.1377** | 0.1797 | 0.092 | 0.1217 |

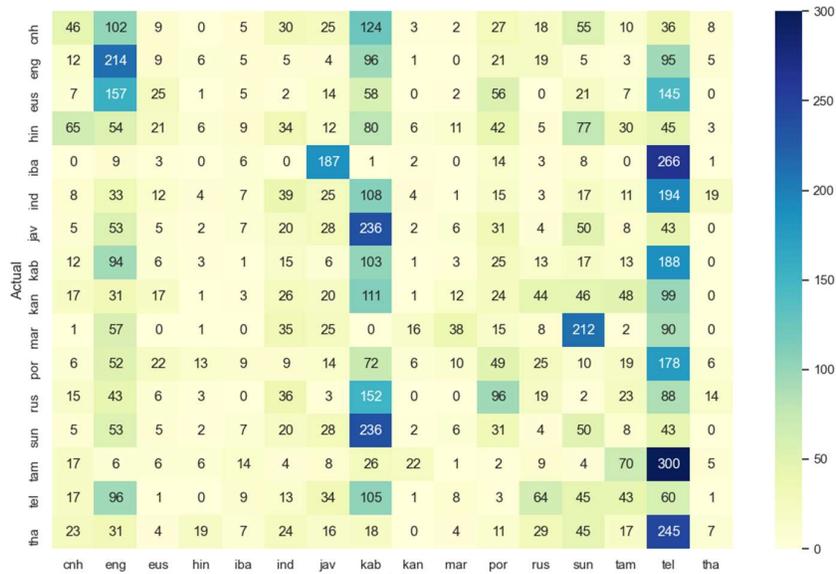

Figure 2: The confusion matrix for the validation set

## Acknowledgments

The authors are grateful to

- colleagues at NTR Labs Machine Learning Research group for the discussions and support;
- Anonymous reviewers who suggested multiple improvements to this memo and helped fix errors

## References

Abdullah, B. M. *et al.* (2020) 'Cross-domain adaptation of spoken language identification for related languages: The curious case of Slavic languages', *arXiv*, (1). doi: 10.21437/interspeech.2020-2930.

Bartz, C. *et al.* (2017) 'Language identification using deep convolutional recurrent neural networks', *Lecture Notes in Computer Science (including subseries Lecture Notes in Artificial Intelligence and Lecture Notes in Bioinformatics)*, 10639 LNCS(1), pp. 880–889. doi: 10.1007/978-3-319-70136-3_93.

Bhattacharya, *et al. (*2017) 'Deep Speaker Embeddings for Short-Duration Speaker Verification', in *Interspeech 2017*. ISCA: ISCA, pp. 1517–1521. doi: 10.21437/Interspeech.2017-1575.

Black A. W., (2019) 'CMU Wilderness Multilingual Speech Dataset,' ICASSP 2019 - 2019 IEEE International Conference on Acoustics, Speech and Signal Processing (ICASSP), pp. 5971-5975, doi: 10.1109/ICASSP.2019.8683536.

Cai, W., et al. (2018) 'Exploring the encoding layer and loss function in end-to-end speaker and language recognition system', arXiv. doi: 10.21437/odyssey.2018-11.

Chowdhury, F. A. R. R. *et al.* (2018) 'Attention-Based Models for Text-Dependent Speaker Verification', in *ICASSP, IEEE International Conference on Acoustics, Speech and Signal Processing - Proceedings*, pp. 5359–5363. doi: 10.1109/ICASSP.2018.8461587.

Dehak, N. *et al.* (2011a) 'Language recognition via i-vectors and dimensionality reduction', in *Proceedings of the Annual Conference of the International Speech Communication Association, INTERSPEECH*, pp. 857–860.

Dehak, N. *et al.* (2011b) 'Front-End Factor Analysis For Speaker Verification', IEEE Transactions On Audio, Speech And Language Processing 1, pp. 1–11. Available at: http://groups.csail.mit.edu/sls//publications/2010/Dehak_IEEE_Transactions.pdf (Accessed: 28 March 2021).

Draghici, A. *et al.* (2020) 'A study on spoken language identification using deep neural networks', *ACM International Conference Proceeding Series*, (July), pp. 253–256. doi: 10.1145/3411109.3411123.

He, F. *et al.* (2020) 'Open-source multi-speaker speech corpora for building gujarati, kannada, malayalam, marathi, tamil and telugu speech synthesis systems', in *LREC 2020 - 12th International Conference on Language Resources and Evaluation, Conference*

Proceedings, pp. 6494–6503. Available at: http://www.openslr.org/78/ (Accessed: 21 April 2021).

Juan S.S. et al. (2014a) 'Semi-Supervised G2p Bootstrapping And Its Application to ASR for a Very Under-Resourced Language : Iban' Grenoble, France. *SLTU-2014*:14–16.

Juan S.S. et al. (2014b) 'Using closely-related language to build an ASR for a very under-resourced language: Iban.' In *Oriental COCOSDA 2014 - 17th Conference of the Oriental Chapter of the International Coordinating Committee on Speech Databases and Speech I/O Systems and Assessment / CASLRE (Conference on Asian Spoken Language Research and Evaluation)*. Institute of Electrical and Electronics Engineers Inc.

Kanagasundaram, A. et al. (2011) 'i-Vector based speaker recognition on short utterances', in *Proceedings of the Annual Conference of the International Speech Communication Association, INTERSPEECH*, pp. 2341–2344. Available at: https://www.researchgate.net/publication/230643046_i-vector_Based_Speaker_Recognition_on_Short_Utterances (Accessed: 28 March 2021).

Kjartansson, O. et al. (2018) 'Crowd-Sourced Speech Corpora for Javanese, Sundanese, Sinhala, Nepali, and Bangladeshi Bengali', in. International Speech Communication Association, pp. 52–55. doi: 10.21437/sltu.2018-11.

Koluguri, N. R. et al. (2020) 'SpeakerNet: 1D Depth-wise Separable Convolutional Network for Text-Independent Speaker Recognition and Verification'. Available at: http://arxiv.org/abs/2010.12653 (Accessed: 20 March 2021).

Kriman, S. et al. (2020) 'Quartznet: Deep Automatic Speech Recognition with 1D Time-Channel Separable Convolutions', in *ICASSP, IEEE International Conference on Acoustics, Speech and Signal Processing - Proceedings*, pp. 6124–6128. doi: 10.1109/ICASSP40776.2020.9053889.

Latif, S. et al. (2020) 'Deep representation learning in speech processing: Challenges, recent advances, and future trends', *arXiv*, pp. 1–25.

Leonard R. and Doddington G., (1974) 'Automatic language identification.' Technical Report RADC-TR74-200 (Air Force Rome Air Development Center, Technical Report) August 1974

Lopez-Moreno, I. et al. (2014) 'Automatic language identification using deep neural networks', *IEEE International Conference on Acoustic, Speech and Signal Processing*.

van der Merwe, R. (2020) 'Triplet Entropy Loss: Improving The Generalisation of Short Speech Language Identification Systems'. Available at: http://arxiv.org/abs/2012.03775.

Navrátil, J. (2006) '*Automatic Language Identification*, *Multilingual Speech Processing*.' doi: 10.1016/b978-012088501-5/50011-1.

Park, D. S. et al. (2019) 'Specaugment: A simple data augmentation method for automatic speech recognition', *Proceedings of the Annual Conference of the International Speech Communication Association, INTERSPEECH*, 2019-Septe, pp. 2613–2617. doi: 10.21437/Interspeech.2019-2680.

Rao, K. and Nandi, D. (2015) Language Identification—A Brief Review. 10.1007/978-3-319-17725-0_2.

Sarthak, et al. (2019) 'Spoken language identification using convNets', *Lecture Notes in Computer Science (including subseries Lecture Notes in Artificial Intelligence and Lecture Notes in Bioinformatics)*, 11912 LNCS, pp. 252–265. doi: 10.1007/978-3-030-34255-5_17.

Snyder, D. et al. (2018) 'X-Vectors: Robust DNN Embeddings for Speaker Recognition', in *ICASSP, IEEE International Conference on Acoustics, Speech and Signal Processing - Proceedings*, pp. 5329–5333. doi: 10.1109/ICASSP.2018.8461375.

Wong, K. E. (2004) 'Automatic Spoken Language Identification Utilizing Acoustic and Phonetic Speech Information', *PhD Thesis*, (Queensland University of Technology).